\documentclass[twocolumn,nopacs,floatfix,amsmath,nofootinbib,amssymb,preprintnumbers,floatfix]{revtex4}
\usepackage{longtable,lscape,booktabs}
\usepackage{txfonts}
\usepackage{overpic}
\usepackage{amssymb}
\usepackage{float}
\usepackage{placeins}
\usepackage{slashed}
\usepackage{color}
\usepackage{longtable,lscape,bm}
\usepackage{multirow}
\usepackage{graphicx,color,dcolumn,bm}
\usepackage[colorlinks,
            citecolor=blue,
            anchorcolor=red,
            menucolor=red,
            linkcolor=red,
            filecolor=red,
            runcolor=red,
            urlcolor=blue,
            frenchlinks=false]{hyperref}

\allowdisplaybreaks
\newcommand{\rev}[1]{{\color{blue}#1}}

\begin{document}

\title{Determination of the $\eta$-$\eta^{\prime}$ Mixing Angle from Transition Form Factors within the Self-consistent Light-front Quark Model}

\author{Shuai Xu$^{1}$} \email{xushuai@zknu.edu.cn}

\affiliation{$^1$School of Physics and Telecommunications Engineering, Zhoukou Normal University, Zhoukou 466001, China\\}

\begin{abstract}
We study the $\eta$-$\eta^{\prime}$ system in the self-consistent light-front quark model (LFQM) with the quark flavor mixing scheme, where the single mixing angle is
constrained by the transition form factors (TFFs) $F_{\eta^{(\prime)}\gamma}(Q^2)$. A continuous combined $\chi^2$ fit to the CLEO and BABAR data gives $\theta=41.55^{+0.57}_{-0.60}\,{}^\circ$, in close agreement with the recent LHCb result $(41.6^{+1.0}_{-1.2})^\circ$. Using this mixing angle, we calculate the decay constants, light-cone distribution amplitudes (LCDAs), Gegenbauer moments, $\xi$-moments, transverse momentum moments, electromagnetic interaction radii and two-photon decay widths of the $\eta$ and $\eta^{\prime}$ mesons. The predicted $a_2^\eta$ and $\langle\xi^2\rangle_\eta$ are consistent with the latest BABAR results, and the electromagnetic interaction radii also agree well with the A2 and BESIII measurements. These results demonstrate that TFF data strongly constrain the light-front description of the nonperturbative properties of the $\eta$-$\eta^{\prime}$ system. After the fit and LFQM-input uncertainties are propagated, the predicted widths are $0.428^{+0.038}_{-0.035}$~keV and $3.067^{+0.190}_{-0.180}$~keV for $\eta\to\gamma\gamma$ and $\eta^{\prime}\to\gamma\gamma$, respectively. The central values are
lower than the corresponding experimental values, indicating that the real-photon point provides a more sensitive test of the absolute TFF normalization.\\
\textbf{Keywords:} Quark Flavor Mixing; Transition Form Factors; Light-Front Quark Model\\
\textbf{PACS:} 12.38.Bx; 12.15.Ff; 12.10.Kt
\end{abstract}

\maketitle

\section{Introduction}\label{Introduction}%
The $\eta$ and $\eta^{\prime}$ mesons have long served as a sensitive probe of nonperturbative QCD. The physical states are not pure quark-antiquark flavor eigenstates, but mixtures of nonstrange and strange components. Their properties are tied to spontaneous chiral symmetry breaking, explicit $SU(3)$ flavor breaking and the axial $U_{A}(1)$ anomaly \cite{Veneziano:1979ec,Witten:1979vv,Leutwyler:1997yr}. The large $\eta^{\prime}$ mass, in particular, lies outside the simple Goldstone boson picture and points directly to flavor-singlet anomaly dynamics. The $\eta$-$\eta^{\prime}$ system therefore remains a useful arena in which to test models of hadron structure.

In the quark flavor basis, the physical $\eta$ and $\eta^{\prime}$ mesons are commonly described by a single mixing angle together with two decay constants for the nonstrange and strange components. This framework was discussed in early work \cite{Schechter:1992iz} and later developed into the widely used Feldmann-Kroll-Stech (FKS) scheme
\cite{Feldmann:1998vh,Feldmann:1998sh,Feldmann:1999uf}. The mixing angle has been studied using phenomenological, analytical and lattice QCD approaches \cite{Bramon:1997va,Cao:1999fs,Escribano:2005qq,Pham:2015ina,Michael:2013gka}. In particular, analyses based on TFFs and two-photon
decays generally favor a quark flavor mixing angle around $40^\circ$ \cite{Kroll:2005sd,Cao:2012nj}. The recent LHCb measurement of $\eta$-$\eta^{\prime}$ mixing in $B^{0}_{(s)}\to J/\psi\eta^{(\prime)}$ decays gives $(41.6^{+1.0}_{-1.2})^{\circ}$ \cite{LHCb:2025sgp}, offering a precise reference value for independent model determinations.

The TFFs $F_{\eta\gamma}(Q^2)$ and $F_{\eta'\gamma}(Q^2)$ provide a sensitive probe of the pseudoscalar-meson structure over a broad range of momentum transfers, with particular sensitivity to the balance between the nonstrange and strange components. Measurements from CELLO \cite{CELLO:1990klc}, CLEO \cite{CLEO:1997fho} and BABAR \cite{BaBar:2011nrp,BaBar:2006ash} therefore provide important constraints on the mixing angle. The TFFs have been studied in several frameworks, including the LFQM \cite{Choi:2017zxn}, the nonlocal NJL model \cite{GomezDumm:2016bxp}, the Bethe-Salpeter/Dyson-Schwinger (BSE/DSE) equations \cite{Ding:2018xwy}, the light-cone sum rules analyses based on updated $\eta^{(\prime)}$ LCDAs \cite{Hu:2026rfj}, and recent dispersive approaches \cite{Holz:2024diw,Holz:2024lom,Messerli:2025rnv}. The LFQM provides a relativistic framework for describing hadronic bound states and exclusive processes, which has been widely applied to meson decay constants, form factors, distribution amplitudes and related observables \cite{Choi:2007yu,Jaus:1999zv,Cheng:2003sm}. The self-consistent formulation based on the replacement $M\to M_{0}$ improves the treatment of covariance and zero-mode contributions in decay constants, distribution amplitudes and weak TFFs \cite{Choi:2013mda,Choi:2017uos,Chang:2018zjq}, and is closely related to the Bakamjian--Thomas construction \cite{Bakamjian:1953kh,Keister:1991sb}. Recently, this framework has been
successfully applied to pseudoscalar electromagnetic form factors, pseudoscalar and vector-meson LCDAs, and quarkonium LCDAs
\cite{Xu:2025ntz,Li:2026wmb,Li:2026wad,Xu:2026zli}, demonstrating that a common light-front wave function can consistently connect different nonperturbative observables. These developments make the self-consistent LFQM a suitable framework for studying the $\eta$-$\eta^{\prime}$ system and its TFFs within a unified description.

In this work, we compute $F_{\eta\gamma}(Q^{2})$ and $F_{\eta^{\prime}\gamma}(Q^{2})$ as functions of the mixing angle $\theta$ and determine it from a continuous combined $\chi^{2}$ fit to the available TFF data. We quote separately the fit uncertainty and the uncertainty propagated from the LFQM inputs. The resulting angle, $\theta=41.55^\circ$, is then used throughout the calculations of the decay constants, LCDAs, their moments, electromagnetic interaction radii, real-photon TFFs, and two-photon decay widths.

The remainder of this paper is organized as follows. In section~\ref{sec:lfqm}, we present the quark flavor mixing scheme and the self-consistent LFQM. In section~\ref{sec:tff}, we calculate the TFFs and determine the $\eta$-$\eta^{\prime}$ mixing angle from a fit to experimental data. In section~\ref{sec:structure}, we discuss the results of decay constants, LCDAs, related moments, electromagnetic interaction radii and two-photon decay widths. Section~\ref{sec:summary} is reserved for a summary.

\section{Theoretical Framework}
\label{sec:lfqm}

\subsection{Quark flavor mixing scheme}

To describe the $\eta$-$\eta^{\prime}$ system, we employ the quark flavor basis
\cite{Feldmann:1998vh,Feldmann:1998sh,Feldmann:1999uf}. The nonstrange and strange flavor states are defined as
\begin{align}
|\eta_q\rangle
=
\frac{1}{\sqrt{2}}
\left(
|u\bar u\rangle+|d\bar d\rangle
\right),~~
|\eta_s\rangle
=
|s\bar s\rangle .
\end{align}

Assuming that the OZI-violating effects are small, the physical $\eta$ and $\eta^{\prime}$ states are expressed in terms of a single mixing angle $\theta$:
\begin{align}
\begin{pmatrix}
|\eta\rangle\\[2pt]
|\eta^{\prime}\rangle
\end{pmatrix}
=
\begin{pmatrix}
\cos\theta & -\sin\theta\\
\sin\theta & \cos\theta
\end{pmatrix}
\begin{pmatrix}
|\eta_q\rangle\\[2pt]
|\eta_s\rangle
\end{pmatrix}.
\end{align}

The axial-vector currents in the quark flavor basis are
\begin{align}
J_{\mu5}^{q}
=
\frac{1}{\sqrt{2}}
\left(
\bar u\gamma_{\mu}\gamma_{5}u
+
\bar d\gamma_{\mu}\gamma_{5}d
\right),~~
J_{\mu5}^{s}
=
\bar s\gamma_{\mu}\gamma_{5}s .
\end{align}
The corresponding decay constants are defined by
\begin{align}
\langle 0|J_{\mu5}^{\rev{f}}|P(p)\rangle
&=
i f_{P}^{f}p_{\mu},
\qquad
f=q,s,
\qquad
P=\eta,\eta^{\prime}.
\end{align}

Within the FKS scheme, they satisfy
\begin{align}
\begin{pmatrix}
f_{\eta}^{q} & f_{\eta}^{s}\\
f_{\eta^{\prime}}^{q} & f_{\eta^{\prime}}^{s}
\end{pmatrix}
=
\begin{pmatrix}
f_q\cos\theta & -f_s\sin\theta\\
f_q\sin\theta & f_s\cos\theta
\end{pmatrix},
\end{align}
where $f_q$ and $f_s$ denote the decay constants of the nonstrange and strange flavor states, respectively, and are calculated from the corresponding light-front wave functions as
\begin{align}
f_f
&=
\frac{\sqrt{2N_c}}{8\pi^3}
\int_0^1\!\mathrm{d}x
\int\!\mathrm{d}^2{\bf k}_\perp
\frac{m_f\,\phi_R^f(x,{\bf k}_\perp)}
{\sqrt{{\bf k}_\perp^2+m_f^2}}.
\label{eq:fflfqm}
\end{align}

Here $\phi_R^f(x,{\bf k}_\perp)$ is introduced in Eq.~\eqref{eq:lfwf-factorization} below.

\subsection{Self-consistent LFQM description and TFFs}

In the quark flavor basis introduced above, the $\eta_q$ and $\eta_s$ components are treated as pseudoscalar quark-antiquark bound states within the self-consistent LFQM.
A pseudoscalar flavor state $P_f$, with four-momentum $p$ and spin projection $S=S_z=0$, is represented by
\begin{align}
|P_f(p)\rangle
=
\sum_{\lambda_1,\lambda_2}
\int
\frac{\mathrm{d}x\,\mathrm{d}^{2}{\bf k}_{\perp}}{16\pi^3}
\Psi_{\lambda_1\lambda_2}^{00,f}
(x,{\bf k}_{\perp})
|f\bar f\rangle ,
\end{align}
where $x$ is the longitudinal momentum fraction carried by the quark, ${\bf k}_{\perp}$ is the relative transverse momentum, and $\lambda_{1,2}$ denote the light-front helicities of the constituent quark and antiquark. Here $q$ denotes the isospin-symmetric nonstrange sector with $m_u=m_d\equiv m_q$.

The light-front wave function is factorized into spin-orbit and radial parts:
\begin{align}
\Psi_{\lambda_1\lambda_2}^{00,f}
(x,{\bf k}_{\perp})
=
S_{\lambda_1\lambda_2}^{00,f}
(x,{\bf k}_{\perp})
\phi_R^f(x,{\bf k}_{\perp}),
\label{eq:lfwf-factorization}
\end{align}
where the spin-orbit wave function $S_{\lambda_1\lambda_2}^{00,f}$ is obtained through the Melosh transformation
\cite{Jaus:1999zv,Cheng:2003sm}.

For the radial wave function, we adopt the Gaussian form commonly used in LFQM analyses
\cite{Choi:2007yu,Choi:2017uos}:
\begin{align}
\phi_R^f(x,{\bf k}_{\perp})
=
\frac{4\pi^{3/4}}{\beta_f^{3/2}}
\sqrt{\frac{\partial k_z}{\partial x}}
\exp\left[
-\frac{{\bf k}_{\perp}^{2}+k_z^{2}}{2\beta_f^{2}}
\right].
\end{align}
Here
\begin{align}
\frac{\partial k_z}{\partial x}
=
\frac{M_{0f}}{4x(1-x)},~~
M_{0f}^{2}
=
\frac{{\bf k}_{\perp}^{2}+m_f^{2}}{x(1-x)}.
\end{align}

The constituent quark masses and Gaussian parameters are taken from our previous studies of pseudoscalar and vector mesons
\cite{Xu:2025ntz,Li:2026wmb,Li:2026wad,Xu:2026zli}
and are summarized in Table~\ref{tab:input}.

\begin{table*}[!t]
\centering
\footnotesize
\caption{Constituent-quark masses and Gaussian parameters
(in GeV). The central values and uncertainties used in the
numerical analysis are taken from Refs.~\cite{Xu:2025ntz,Li:2026wmb,Li:2026wad,Xu:2026zli}.}
\label{tab:input}

\setlength{\tabcolsep}{45pt}
\renewcommand{\arraystretch}{1.2}
\begin{tabular}{cccc}
\hline\hline
$m_q$ & $m_s$ & $\beta_{q\bar q}$ & $\beta_{s\bar s}$ \\
\hline
$0.25^{+0.01}_{-0.01}$ &
$0.50^{+0.02}_{-0.02}$ &
$0.321^{+0.016}_{-0.016}$ &
$0.348^{+0.006}_{-0.006}$ \\
\hline\hline
\end{tabular}
\end{table*}

\begin{table*}[!htb]
\centering
\footnotesize
\caption{Combined $\chi^2$ profile near the minimum. Here $\Delta\chi^2=\chi^2-\chi^2_{\min}$, with $\chi^2_{\min}=32.284$. The best-fit angle is obtained from the continuous profile rather than from the tabulated grid.}
\label{tab:chi2}
\setlength{\tabcolsep}{40pt}
\renewcommand{\arraystretch}{1.05}
\begin{tabular}{ccccc}
\hline\hline
$\theta$ & $\chi^2_\eta$ & $\chi^2_{\eta^{\prime}}$ & $\chi^2$ & $\Delta\chi^2$\\
\hline
$41.0^{\circ}$ & 15.229 & 18.913 & 34.142 & 1.859\\
$41.1^{\circ}$ & 15.244 & 18.285 & 33.530 & 1.246\\
$41.2^{\circ}$ & 15.340 & 17.699 & 33.039 & 0.756\\
$41.3^{\circ}$ & 15.515 & 17.155 & 32.671 & 0.387\\
$41.4^{\circ}$ & 15.771 & 16.653 & 32.424 & 0.141\\
$41.5^{\circ}$ & 16.108 & 16.192 & 32.300 & 0.017\\
$41.6^{\circ}$ & 16.526 & 15.771 & 32.298 & 0.014\\
$41.7^{\circ}$ & 17.025 & 15.391 & 32.417 & 0.133\\
$41.8^{\circ}$ & 17.606 & 15.052 & 32.658 & 0.375\\
$41.9^{\circ}$ & 18.269 & 14.752 & 33.021 & 0.737\\
$42.0^{\circ}$ & 19.014 & 14.492 & 33.506 & 1.222\\
\hline\hline
\end{tabular}
\end{table*}
Using the light-front wave functions, we calculate the TFFs for
$P(p)\rightarrow\gamma^{*}(q)\gamma(p-q)$. The transition matrix element of the electromagnetic current is defined as
\begin{align}
\Gamma^{\mu}
=
\langle
\gamma(p-q,\epsilon)
|
J_{\rm em}^{\mu}(0)
|
P(p)
\rangle
=
i e^{2}F_{P\gamma}(Q^{2})
\epsilon^{\mu\nu\rho\sigma}
\epsilon_{\nu}^{*}
p_{\rho}q_{\sigma}.
\end{align}
Here $Q^{2}=-q^{2}={\bf q}_{\perp}^{2}$ is the spacelike momentum transfer and $\epsilon_{\nu}$ is the polarization vector of the on-shell photon with momentum $p-q$.

Within the self-consistent LFQM, the light-front loop integral associated with a quark of flavor $f$ is given by
\cite{Choi:2017zxn}
\begin{align}
I_{\rm tot}^{m_f}(Q^{2})
=
\frac{\sqrt{2N_c}}{4\pi^{3}}
\int_{0}^{1}\frac{\mathrm{d}x}{1-x}
\int \mathrm{d}^{2}{\bf k}_{\perp}
\frac{m_f}{M_{0f}^{2}+Q^{2}}
\frac{
\phi_R^f(x,{\bf k}_{\perp})
}{
\sqrt{{\bf k}_{\perp}^{2}+m_f^{2}}
}.
\end{align}

The TFFs of the nonstrange and strange flavor components are therefore
\begin{align}
F_{q\gamma}(Q^{2})
=
\frac{e_u^{2}+e_d^{2}}{\sqrt{2}}\,
I_{\rm tot}^{m_q}(Q^{2}),~~
F_{s\gamma}(Q^{2})
=
e_s^{2}\,
I_{\rm tot}^{m_s}(Q^{2}),
\end{align}
respectively.

Combining these LFQM results with the quark flavor mixing relations, the TFFs are
\begin{align}
F_{\eta\gamma}(Q^{2})
&=
\cos\theta\,F_{q\gamma}(Q^{2})
-
\sin\theta\,F_{s\gamma}(Q^{2})
\nonumber\\
&=
\cos\theta\,
\frac{e_u^{2}+e_d^{2}}{\sqrt{2}}\,
I_{\rm tot}^{m_q}(Q^{2})-
\sin\theta\,
e_s^{2}I_{\rm tot}^{m_s}(Q^{2}),
\end{align}
and
\begin{align}
F_{\eta^{\prime}\gamma}(Q^{2})
&=
\sin\theta\,F_{q\gamma}(Q^{2})
+
\cos\theta\,F_{s\gamma}(Q^{2})
\nonumber\\
&=
\sin\theta\,
\frac{e_u^{2}+e_d^{2}}{\sqrt{2}}\,
I_{\rm tot}^{m_q}(Q^{2})+
\cos\theta\,
e_s^{2}I_{\rm tot}^{m_s}(Q^{2}).
\end{align}

Thus, the self-consistent light-front wave functions determine both the nonstrange and strange contributions, while the mixing angle $\theta$ specifies how these two flavor components combine to form the physical $\eta$ and $\eta^{\prime}$ TFFs. Once the constituent quark masses and Gaussian parameters are specified, $\theta$ is the only remaining parameter and can be constrained by the measured TFF data.

\section{Combined $\chi^{2}$ analysis and extraction of the mixing angle}
\label{sec:tff}

In the numerical analysis, we use the TFF data measured by the CLEO
\cite{CLEO:1997fho} and BABAR \cite{BaBar:2011nrp} collaborations, covering the region
$2~\mathrm{GeV}^{2}\leq Q^{2}<40~\mathrm{GeV}^{2}$. The quantity compared with the theoretical prediction at each data point is $Q^{2}F_{P\gamma}(Q^{2})$.

For each channel $a=\eta,\eta^{\prime}$, we define

\begin{align}
\chi_a^2(\theta)
&=\sum_{i=1}^{N_a}
\left[
\frac{[Q_i^2F_{a\gamma}(Q_i^2)]_{\rm exp}
-[Q_i^2F_{a\gamma}(Q_i^2;\theta)]_{\rm th}}
{\delta_{a,i}}
\right]^2,
\\
\chi^2(\theta)
&=\chi_\eta^2(\theta)+\chi_{\eta^{\prime}}^2(\theta).
\label{eq:combinedchi2}
\end{align}
where $\delta_{a,i}$ denotes the experimental uncertainty of the $i$th data point. The combined sample contains 32 measurements and one fitted parameter, so the goodness of fit is evaluated with 31 degrees of freedom. Table~\ref{tab:chi2} lists the profile near its minimum. A continuous minimization gives
\begin{align}
\theta_{\rm best}&=41.55^\circ,
&\chi^2_{\min}&=32.284,
\nonumber\\
\chi^2_{\min}/31&=1.041,
&p&=0.403.
\label{eq:centralfit}
\end{align}

For one fitted parameter, the $68.3\%$ confidence interval is determined by $\chi^2(\theta)-\chi^2_{\min}=1$, which gives $\Delta\theta_{\rm fit}
={}^{+0.41}_{-0.40}\,{}^\circ$. Propagating the uncertainties of the LFQM inputs in Table~\ref{tab:input} through the complete combined fit gives $\Delta\theta_{\rm par}
={}^{+0.41}_{-0.44}\,{}^\circ$.

\begin{figure*}[!htbp]
\centering
\includegraphics[width=13cm,height=11cm]{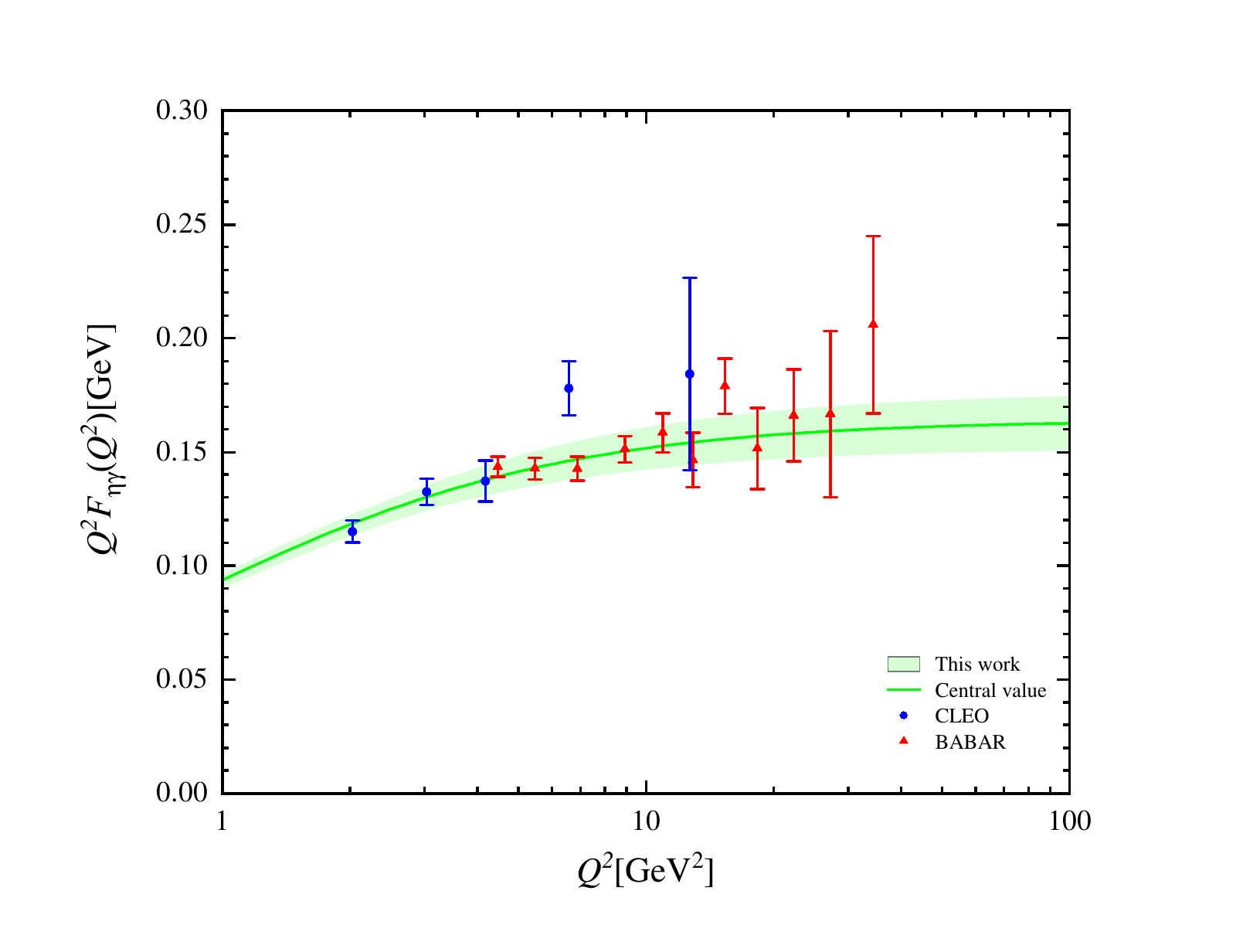}
\includegraphics[width=13cm,height=11cm]{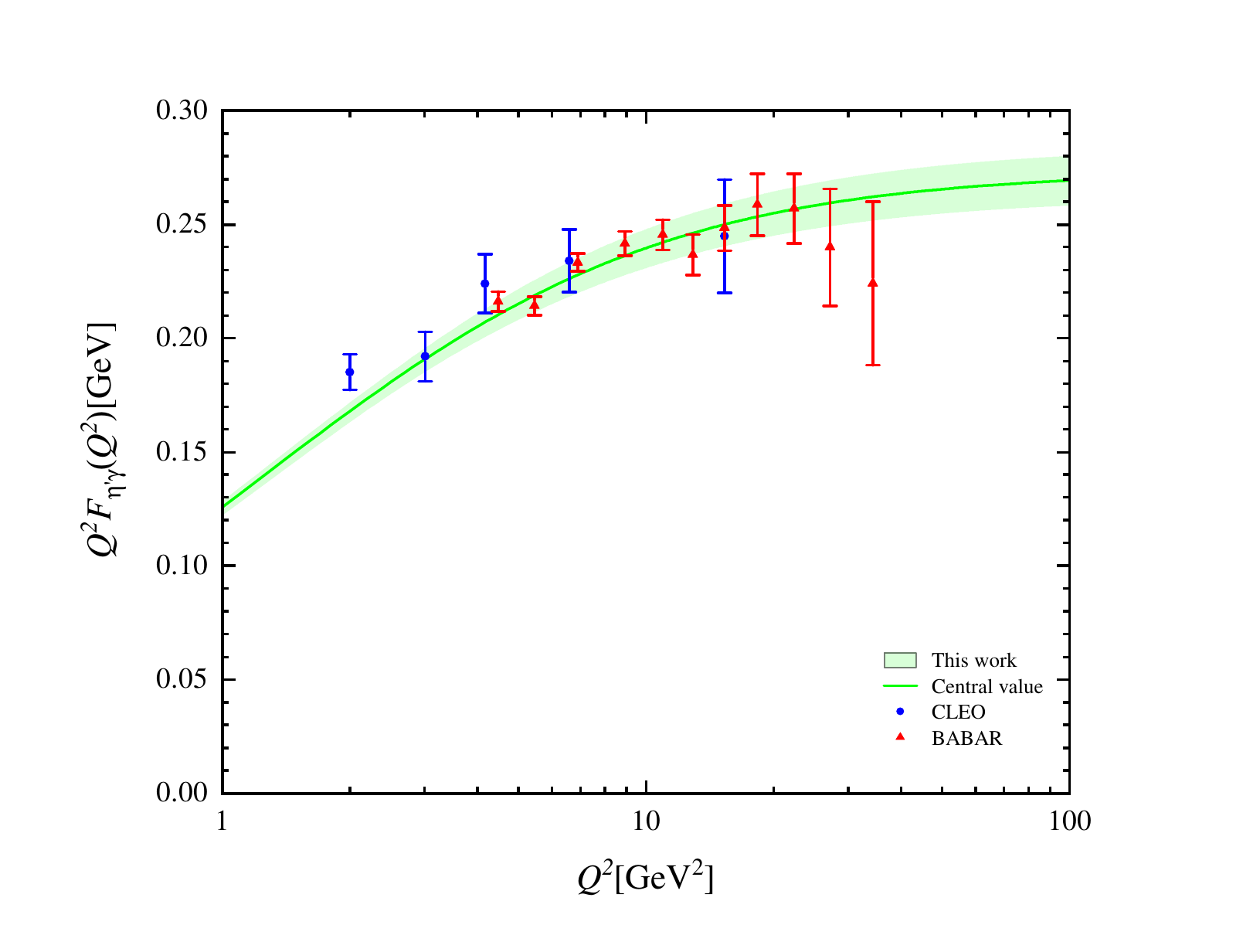}
\caption{Transition form factors obtained in the self-consistent LFQM with the best-fit angle $\theta=41.55^{\circ}$. The upper panel shows $Q^{2}F_{\eta\gamma}(Q^{2})$ and the lower panel shows $Q^{2}F_{\eta^{\prime}\gamma}(Q^{2})$. The shaded bands include the propagated fit and LFQM-input uncertainties. The experimental data are taken from CLEO~\cite{CLEO:1997fho} and BABAR~\cite{BaBar:2011nrp}.}
\label{fig:tff_pair}
\end{figure*}

The result with the two uncertainty components displayed separately is

\begin{align}
\theta
=41.55^{+0.41}_{-0.40}\big|_{\rm fit}
{}^{+0.41}_{-0.44}\big|_{\rm par}\,{}^\circ.
\label{eq:theta_components}
\end{align}
Combining the two independent uncertainties in quadrature using their unrounded values, we obtain

\begin{align}
\theta
=41.55^{+0.57}_{-0.60}\,{}^\circ.
\label{eq:thetafinal}
\end{align}
This result agrees with the LHCb determination $(41.6^{+1.0}_{-1.2})^\circ$~\cite{LHCb:2025sgp}. The corresponding TFFs are shown in Fig.~\ref{fig:tff_pair}.

\begin{figure*}[!htbp]
\centering
\includegraphics[width=13cm,height=11cm]{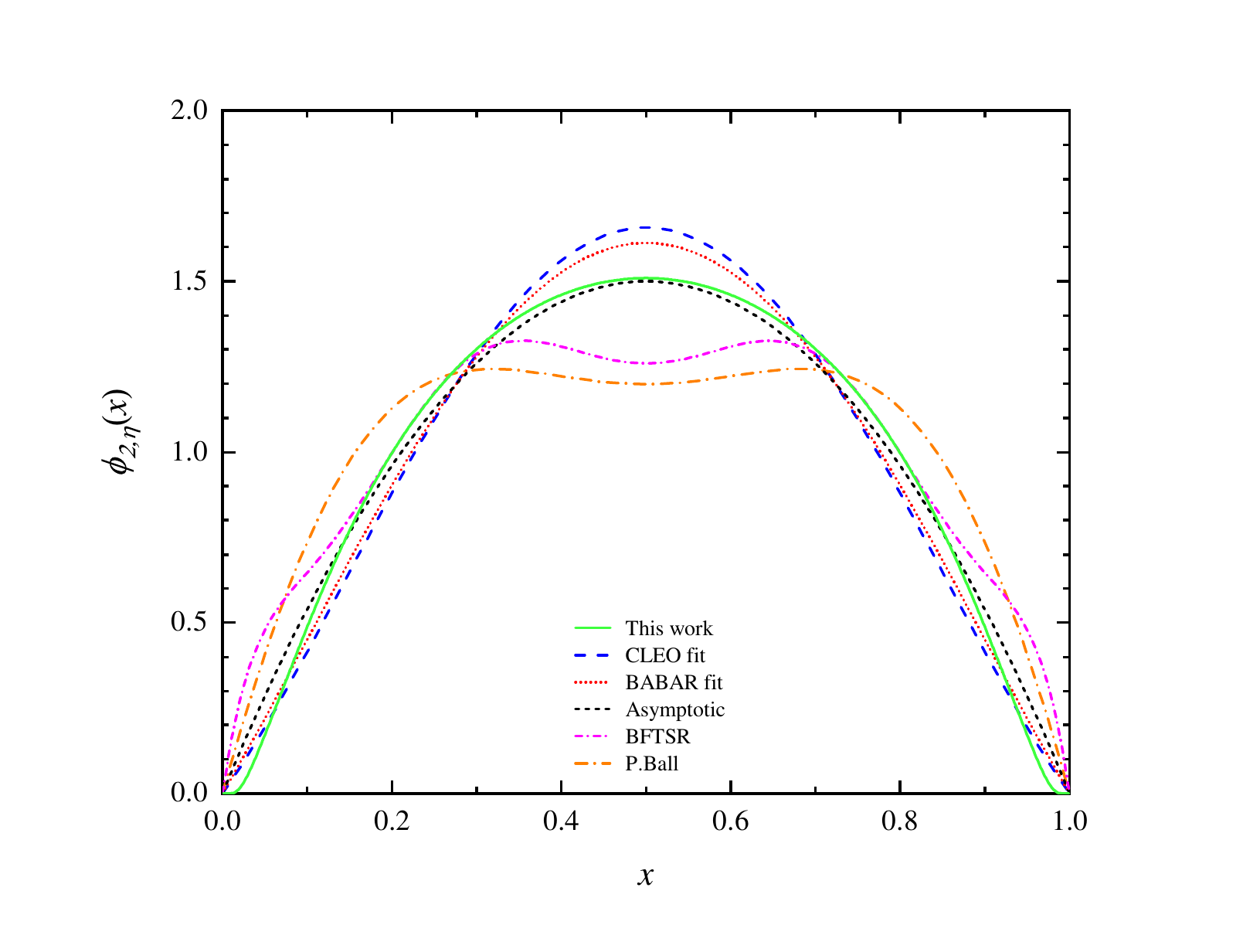}
\includegraphics[width=13cm,height=11cm]{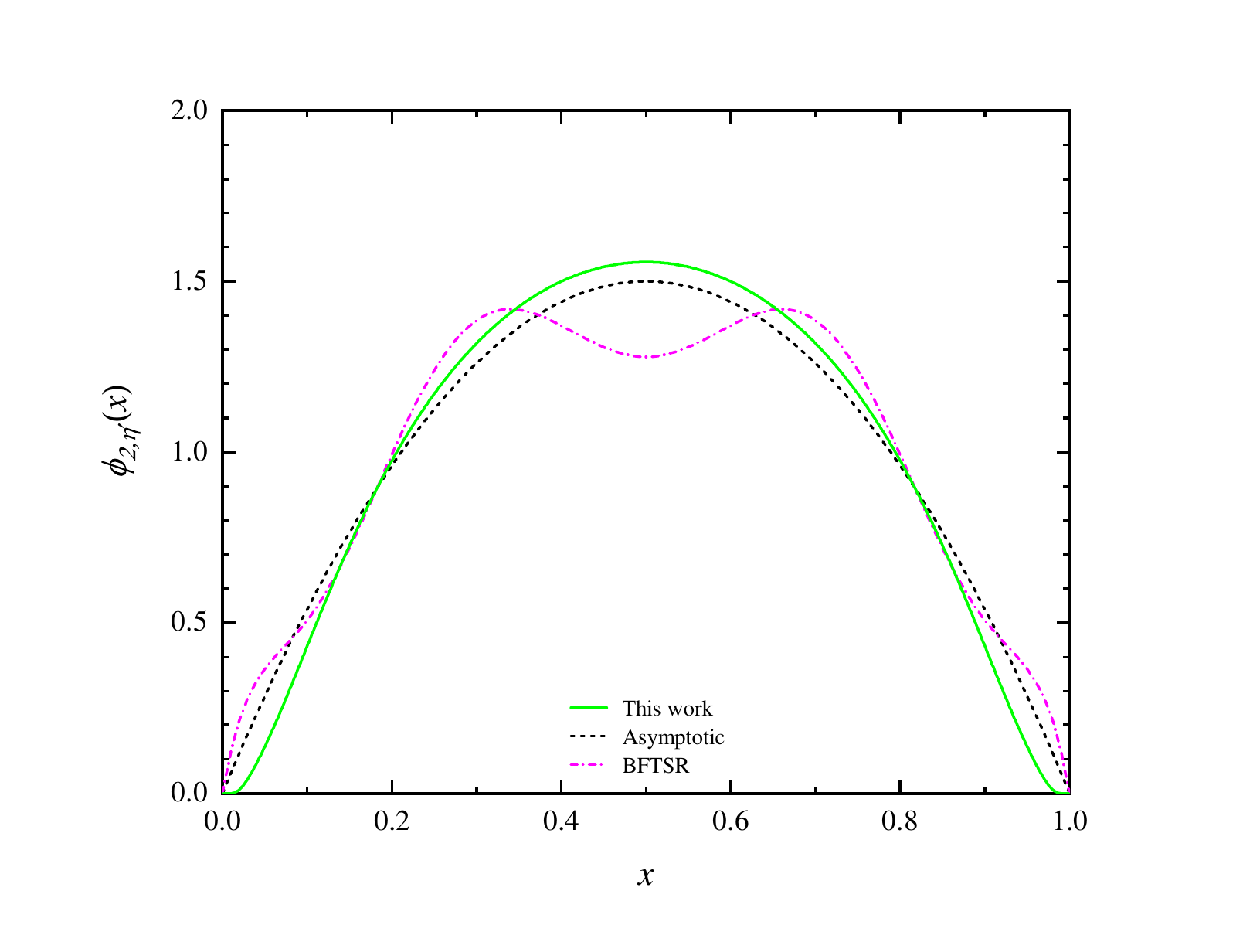}
\caption{Leading-twist LCDAs obtained with $\theta=41.55^{\circ}$. The upper panel shows the $\eta$ LCDA, compared with the asymptotic form and other determinations~\cite{CLEO:1997fho,BaBar:2011nrp,Hu:2021zmy,Ball:2004ye}. The lower panel shows the $\eta^{\prime}$ LCDA, compared with the asymptotic form and the BFTSR prediction~\cite{Hu:2021zmy}.}
\label{fig:lcda_pair}
\end{figure*}

\section{Internal structure of the $\eta$-$\eta^{\prime}$ system}
\label{sec:structure}
Unless otherwise stated, the uncertainties of our results quoted below
combine the mixing-angle fit uncertainty and the LFQM input-parameter
uncertainties in quadrature. For each input-parameter variation,
the mixing angle is refitted before the corresponding observable
is recalculated. Since $f_q$ and $f_s$ are independent of the
mixing angle, their uncertainties arise solely from the LFQM
inputs.
\subsection{Decay constants}

The flavor-basis decay constants are calculated directly from Eq.~\eqref{eq:fflfqm}. Using the parameters in Table~\ref{tab:input}, we obtain
\begin{align}
f_{q}
=
0.131^{+0.004}_{-0.004}~\mathrm{GeV},~~
f_{s}
=
0.174^{+0.003}_{-0.003}~\mathrm{GeV}.
\end{align}
The corresponding FKS values are
\cite{Feldmann:1998vh,Feldmann:1998sh,Feldmann:1999uf}
\begin{align}
f^{\rm FKS}_{q}
&=
(1.07^{+0.02}_{-0.02})f_{\pi}
=
0.139^{+0.003}_{-0.003}~\mathrm{GeV},
\nonumber\\
f^{\rm FKS}_{s}
&=
(1.34^{+0.06}_{-0.06})f_{\pi}
=
0.174^{+0.008}_{-0.008}~\mathrm{GeV},
\end{align}
where $f_{\pi}=0.130~\mathrm{GeV}$ is adopted.

In the isospin-symmetric nonstrange sector with
$m_u=m_d$, $f_q$ coincides with the model value of $f_\pi$.
It is therefore close to $f_\pi$ and slightly smaller than the
FKS value. The strange decay constant agrees with the FKS
determination, with its larger value reflecting $SU(3)$ flavor breaking.

\subsection{Light-cone distribution amplitudes}

The LCDAs describe the longitudinal momentum fraction distributions of quarks at small transverse separation and provide important nonperturbative inputs for hard exclusive processes
\cite{Lepage:1980fj,Chernyak:1983ej}. In the LFQM, the leading-twist LCDA is obtained by integrating the light-front wave function over ${\bf k}_{\perp}$:
\begin{align}
\phi_{P}(x)
=
\frac{\sqrt{2N_{c}}}{8\pi^{3}f_{P}}
\int \mathrm{d}^{2}{\bf k}_{\perp}
\frac{
m_{f}\,\rev{\phi_{R}^{f}}(x,{\bf k}_{\perp})
}{
\sqrt{{\bf k}_{\perp}^{2}+m_{f}^{2}}
}.
\end{align}

The $\eta$ and $\eta'$ LCDAs are constructed from the nonstrange and strange components with the mixing angle $\theta=41.55^\circ$. The distributions are shown in Fig.~\ref{fig:lcda_pair} together with the asymptotic form $\phi_{\rm as}(x)=6x(1-x)$. Both LCDAs are symmetric around $x=1/2$, reflecting the symmetric structure of the corresponding flavor components. Compared with the asymptotic distribution, the LCDAs are slightly narrower with suppressed endpoint regions, due to the finite constituent masses and the Gaussian wave-function parameters in the LFQM. The difference between the $\eta$ and $\eta'$ LCDAs mainly arises from $SU(3)$ flavor breaking and their different quark contents. The longitudinal and transverse moments discussed below further quantify these differences.

\subsection{Gegenbauer moments, $\xi$-moments and transverse momentum moments}
\label{sec:moments}

LCDAs are commonly expanded in terms of Gegenbauer polynomials, with the expansion coefficients identified as the Gegenbauer moments. For the $\eta$ and $\eta^{\prime}$ mesons, the leading-twist LCDAs are expanded as
\begin{align}
\phi_{P}(x,\mu)
&=
\phi_{\rm as}(x)
\sum_{n=0}^{\infty}
a^P_n(\mu)
C_n^{3/2}(2x-1),
\\
a_n^P(\mu)
&=
\frac{4n+6}{3n^2+9n+6}
\int_0^1 \mathrm{d}x\,
C_n^{3/2}(2x-1)\,
\phi_{P}(x,\mu),
\end{align}
where $C^{3/2}_{n}$ are Gegenbauer polynomials. The moments $a^{P}_{n}$ quantify deviations from the asymptotic form.

To characterize the longitudinal and transverse structure of the bound states, we introduce the $\xi$-moments and transverse momentum moments. The $\xi$-moments characterize the longitudinal momentum distribution and are defined as
\begin{align}
\langle\xi^{n}\rangle_{P}
&=
\int_{0}^{1}
\mathrm{d}x\,
\xi^{n}\phi_{P}(x),
\qquad
\xi=2x-1.
\end{align}
The transverse momentum moments characterize the intrinsic transverse motion of the constituent quarks and are defined as
\begin{align}
\langle {\bf k}_{\perp}^{n}\rangle
&=
\int_{0}^{1}
\mathrm{d}x\,
{\bf k}_{\perp}^{n}\phi_{P}(x).
\end{align}
The numerical results for the Gegenbauer and $\xi$-moments are summarized in Table~\ref{tab:moments}. The transverse momentum moments are listed in Table~\ref{tab:kperp}. From these results, one can make the following observations.

\begin{table*}[!t]
\scriptsize
\caption{Gegenbauer moments and $\xi$-moments of the $\eta$ and
$\eta^{\prime}$ mesons.}
\label{tab:moments}
\setlength{\tabcolsep}{15pt}
\renewcommand{\arraystretch}{1.5}
\begin{tabular}{lcccccc}
\hline\hline
& $a^{\eta}_{2}$ & $a^{\eta}_{4}$ & $a^{\eta}_{6}$
& $a^{\eta^{\prime}}_{2}$ & $a^{\eta^{\prime}}_{4}$
& $a^{\eta^{\prime}}_{6}$\\
\hline
This work
& $-0.035^{+0.012}_{-0.013}$
& $-0.035^{+0.004}_{-0.004}$
& $-0.012^{+0.001}_{-0.001}$
& $-0.059^{+0.012}_{-0.012}$
& $-0.036^{+0.004}_{-0.004}$
& $-0.009^{+0.002}_{-0.002}$\\
CLEO fit \cite{CLEO:1997fho}
& $-0.07^{+0.03}_{-0.03}$ & -- & -- & -- & -- & --\\
BABAR fit \cite{BaBar:2011nrp}
& $-0.05^{+0.02}_{-0.02}$ & -- & -- & -- & -- & --\\
BFTSR \cite{Hu:2021zmy}
& $0.090^{+0.031}_{-0.037}$
& $0.025^{+0.003}_{-0.010}$
& $0.033^{+0.055}_{-0.058}$
& $0.033^{+0.042}_{-0.050}$
& $-0.002^{+0.007}_{-0.016}$
& $0.043^{+0.067}_{-0.072}$\\
P. Kroll \cite{Kroll:2012gsh}
& $-0.05^{+0.02}_{-0.02}$ & -- & -- & -- & -- & --\\
SR fit \cite{Agaev:2014wna}
& $0.25^{+0.15}_{-0.15}$ & -- & -- & -- & -- & --\\
P. Ball \cite{Ball:2004ye}
& $0.115$ & $-0.015$ & -- & -- & -- & --\\
\hline
& $\langle\xi^{2}\rangle_{\eta}$
& $\langle\xi^{4}\rangle_{\eta}$
& $\langle\xi^{6}\rangle_{\eta}$
& $\langle\xi^{2}\rangle_{\eta^{\prime}}$
& $\langle\xi^{4}\rangle_{\eta^{\prime}}$
& $\langle\xi^{6}\rangle_{\eta^{\prime}}$\\
\hline
This work
& $0.188^{+0.004}_{-0.004}$
& $0.074^{+0.003}_{-0.003}$
& $0.038^{+0.002}_{-0.002}$
& $0.180^{+0.004}_{-0.004}$
& $0.069^{+0.003}_{-0.003}$
& $0.034^{+0.002}_{-0.002}$\\
CLEO fit \cite{CLEO:1997fho}
& $0.176^{+0.010}_{-0.010}$ & -- & -- & -- & -- & --\\
BABAR fit \cite{BaBar:2011nrp}
& $0.183^{+0.007}_{-0.007}$ & -- & -- & -- & -- & --\\
BFTSR \cite{Hu:2021zmy}
& $0.231^{+0.010}_{-0.013}$
& $0.109^{+0.007}_{-0.007}$
& $0.066^{+0.006}_{-0.006}$
& $0.211^{+0.015}_{-0.017}$
& $0.093^{+0.009}_{-0.009}$
& $0.054^{+0.008}_{-0.008}$\\
P. Kroll \cite{Kroll:2012gsh}
& $0.183^{+0.007}_{-0.007}$ & -- & -- & -- & -- & --\\
SR fit \cite{Agaev:2014wna}
& $0.286^{+0.051}_{-0.051}$ & -- & -- & -- & -- & --\\
P. Ball \cite{Ball:2004ye}
& $0.239$ & $0.110$ & -- & -- & -- & --\\
\hline\hline
\end{tabular}
\end{table*}

\begin{table*}[!t]
\scriptsize
\caption{Transverse-momentum moments (in GeV) of the $\eta$ and
$\eta^{\prime}$ mesons.}
\label{tab:kperp}
\setlength{\tabcolsep}{15pt}
\renewcommand{\arraystretch}{1.5}
\begin{tabular}{lcccccc}
\hline\hline
Meson
& $\langle {\bf k}_{\perp}\rangle$
& $\sqrt{\langle {\bf k}_{\perp}^{2}\rangle}$
& $\sqrt[3]{\langle {\bf k}_{\perp}^{3}\rangle}$
& $\sqrt[4]{\langle {\bf k}_{\perp}^{4}\rangle}$
& $\sqrt[5]{\langle {\bf k}_{\perp}^{5}\rangle}$
& $\sqrt[6]{\langle {\bf k}_{\perp}^{6}\rangle}$\\
\hline
$\eta$
& $0.349^{+0.009}_{-0.009}$
& $0.400^{+0.010}_{-0.010}$
& $0.447^{+0.010}_{-0.010}$
& $0.489^{+0.0110}_{-0.011}$
& $0.529^{+0.012}_{-0.011}$
& $0.565^{+0.012}_{-0.011}$\\
$\eta^{\prime}$
& $0.356^{+0.007}_{-0.007}$
& $0.408^{+0.008}_{-0.008}$
& $0.454^{+0.009}_{-0.008}$
& $0.497^{+0.009}_{-0.009}$
& $0.536^{+0.010}_{-0.009}$
& $0.573^{+0.010}_{-0.009}$\\
\hline\hline
\end{tabular}
\end{table*}

\begin{itemize}
\item
For the $\eta$ meson, we obtain
$a^{\eta}_{2}=-0.035^{+0.012}_{-0.013}$,
$a^{\eta}_{4}=-0.035^{+0.004}_{-0.004}$, and
$a^{\eta}_{6}=-0.012^{+0.001}_{-0.001}$,
while for the $\eta^{\prime}$ meson,
$a^{\eta^{\prime}}_{2}=-0.059^{+0.012}_{-0.012}$,
$a^{\eta^{\prime}}_{4}=-0.036^{+0.004}_{-0.004}$, and
$a^{\eta^{\prime}}_{6}=-0.009^{+0.002}_{-0.002}$.
The second Gegenbauer moments are negative for both mesons, consistent with LCDAs that are slightly narrower than the asymptotic form, as seen in Fig.~\ref{fig:lcda_pair}. Our result for $a^{\eta}_{2}$ lies within the ranges reported by BABAR and the Kroll analysis
\cite{BaBar:2011nrp,Kroll:2012gsh}.

\item
The second $\xi$-moment,
$\langle\xi^{2}\rangle_{\eta}=0.188^{+0.004}_{-0.004}$,
is close to the BABAR and Kroll fit value
$0.183^{+0.007}_{-0.007}$
\cite{BaBar:2011nrp,Kroll:2012gsh},
which supports the longitudinal distribution shape predicted in this work.
The $\xi$-moments decrease with increasing order, reflecting endpoint suppression and the concentration of the distributions near $x=1/2$.

\item
The transverse momentum moments listed in Table~\ref{tab:kperp} increase with the order of the moment, as expected for a smooth positive distribution.
The $\eta^{\prime}$ values are about two percent larger than the $\eta$ values.
This reflects the larger strange component of the $\eta'$, where the heavier strange quark leads to a more localized distribution.
The behavior is consistent with the Heisenberg uncertainty principle and previous findings on the mass dependence of quarkonium properties
\cite{Xu:2026zli}.
\end{itemize}

\subsection{Electromagnetic properties}

The TFFs also allow us to extract electromagnetic quantities at the real photon point and in the low-$Q^2$ region, providing further tests of the internal structure of the mesons. The electromagnetic interaction radii are sensitive to the low-$Q^2$ slope of $F_{P\gamma}(Q^2)$, whereas the decay widths are determined by the value of $F_{P\gamma}(0)$ at $Q^2=0$. These observables provide complementary information on the meson structure.

For neutral pseudoscalar mesons such as $\eta$ and $\eta^{\prime}$, the radius extracted from the TFF slope is an electromagnetic interaction radius, which is defined as
\begin{align}
r^{2}_{P}
&=
-\frac{6}{F_{P\gamma}(0)}
\left.
\frac{
\mathrm{d}F_{P\gamma}(Q^{2})
}{
\mathrm{d}Q^{2}
}
\right|_{Q^{2}=0}.
\end{align}

\begin{table*}[!t]
\centering
\footnotesize
\caption{Comparison of the electromagnetic interaction radii (in fm).}
\label{tab:radii}
\setlength{\tabcolsep}{40pt}
\renewcommand{\arraystretch}{1.5}
\begin{tabular}{lccc}
\hline\hline
& $r_{\eta}$ & $r_{\eta^{\prime}}$ & $r_{\eta}/r_{\eta^{\prime}}$\\
\hline
This work
& $0.666^{+0.022}_{-0.020}$
& $0.610^{+0.020}_{-0.019}$
& $1.091^{+0.002}_{-0.002}$\\
BSE/DSE \cite{Ding:2018xwy}
& $0.83^{+0.40}_{-0.22}$
& $0.73^{+0.34}_{-0.19}$
& $1.14^{+0.01}_{-0.01}$\\
A2, BESIII \cite{A2:2013wad,BESIII:2015zpz}
& $0.67^{+0.03}_{-0.03}$
& $0.61^{+0.03}_{-0.03}$
& $1.10^{+0.07}_{-0.07}$\\
\hline\hline
\end{tabular}
\end{table*}

\begin{table*}[!htb]
\centering
\footnotesize
\caption{$F_{P\gamma}(0)$ and two-photon decay widths.}
\label{tab:widths}
\setlength{\tabcolsep}{27pt}
\renewcommand{\arraystretch}{1.5}

\begin{tabular}{lccc}
\hline\hline
Quantity
& This work
& BSE/DSE \cite{Ding:2018xwy}
& Exp. \cite{ParticleDataGroup:2024cfk}\\
\hline
$F_{\eta\gamma}(0)~(\mathrm{GeV}^{-1})$
& $0.250^{+0.010}_{-0.010}$
& --
& $0.274^{+0.005}_{-0.005}$\\
$F_{\eta^{\prime}\gamma}(0)~(\mathrm{GeV}^{-1})$
& $0.289^{+0.009}_{-0.009}$
& --
& $0.344^{+0.006}_{-0.006}$\\
$\Gamma(\eta\to\gamma\gamma)~(\mathrm{keV})$
& $0.428^{+0.038}_{-0.035}$
& $0.42$
& $0.516^{+0.018}_{-0.018}$\\
$\Gamma(\eta^{\prime}\to\gamma\gamma)~(\mathrm{keV})$
& $3.067^{+0.190}_{-0.180}$
& $4.66$
& $4.35^{+0.14}_{-0.14}$\\
\hline\hline
\end{tabular}
\end{table*}
The results are summarized in Table~\ref{tab:radii} and compared with the continuum BSE/DSE result of ref.~\cite{Ding:2018xwy} and with values extracted from A2 and BESIII measurements
\cite{A2:2013wad,BESIII:2015zpz}. The smaller $\eta^{\prime}$ radius follows from its larger strange quark component and the corresponding more compact distribution. Our predictions agree with the A2 extraction $r_\eta=0.67^{+0.03}_{-0.03}\,\mathrm{fm}$~\cite{A2:2013wad} and the BESIII determination $r_{\eta^{\prime}}=0.61^{+0.03}_{-0.03}\,\mathrm{fm}$~\cite{BESIII:2015zpz}. The ratio $r_{\eta}/r_{\eta^{\prime}}=1.091^{+0.002}_{-0.002}$ is also consistent with the experimental value.

For $\eta$ and $\eta^{\prime}$, the two-photon decay width is
\begin{align}
\Gamma(P\to\gamma\gamma)
&=
\frac{\pi}{4}
\alpha_{\rm em}^{2}
M_{P}^{3}
|F_{P\gamma}(0)|^{2}.
\end{align}

The propagated results for $F_{P\gamma}(0)$ and $\Gamma(P\to\gamma\gamma)$ are given in Table~\ref{tab:widths}. The predicted two-photon decay widths are lower than the corresponding experimental values. Since the finite-$Q^2$ data do not directly constrain $F_{P\gamma}(0)$, the real-photon point provides a more sensitive test of the absolute TFF normalization.
\section{Summary}\label{sec:summary}
We have studied the $\eta$--$\eta^{\prime}$ system in the self-consistent LFQM. A continuous combined fit to the CLEO and BABAR TFF data yields $\theta=41.55^{+0.57}_{-0.60}\,{}^\circ$, in agreement with the independent LHCb value.
Decay constants in the flavor basis are obtained directly from the LFQM wave functions, with $f_q$ coinciding with the model value of $f_\pi$ in the isospin-symmetric nonstrange sector and $f_s$ agreeing with the FKS value. Within the same framework, the predicted LCDAs are symmetric and slightly narrower than the asymptotic form, while their Gegenbauer and $\xi$ moments are consistent with the BABAR and Kroll analyses. The resulting radii, $r_{\eta}=0.666^{+0.022}_{-0.020}$~fm and $r_{\eta^{\prime}}=0.610^{+0.020}_{-0.019}$~fm, agree with the A2 and BESIII extractions, respectively. After full uncertainty propagation, the predicted two-photon decay widths are $\Gamma(\eta\to\gamma\gamma)=0.428^{+0.038}_{-0.035}$~keV and $\Gamma(\eta^{\prime}\to\gamma\gamma)=3.067^{+0.190}_{-0.180}$~keV, both lower than the corresponding experimental values. Overall, the finite-$Q^2$ TFFs, LCDA moments, and interaction radii are consistently described, whereas the real-photon point provides a more sensitive test of the absolute TFF normalization.

\section*{ACKNOWLEDGMENTS}
We are grateful to Prof.~Xing-Gang Wu for valuable discussions and constructive comments. This work is supported by the Scientific Research Project of Colleges and Universities in Henan Province (No. 27AQ140025).


\begin{thebibliography}{99}
\bibitem{Veneziano:1979ec}
G.~Veneziano,
Nucl. Phys. B \textbf{159}, 213-224 (1979)

\bibitem{Witten:1979vv}
E.~Witten,
Nucl. Phys. B \textbf{156}, 269-283 (1979)

\bibitem{Leutwyler:1997yr}
H.~Leutwyler,
Nucl. Phys. B Proc. Suppl. \textbf{64}, 223-231 (1998)

\bibitem{Schechter:1992iz}
J.~Schechter, A.~Subbaraman and H.~Weigel,
Phys. Rev. D \textbf{48}, 339-355 (1993)

\bibitem{Feldmann:1998vh}
T.~Feldmann, P.~Kroll and B.~Stech,
Phys. Rev. D \textbf{58}, 114006 (1998)

\bibitem{Feldmann:1998sh}
T.~Feldmann, P.~Kroll and B.~Stech,
Phys. Lett. B \textbf{449}, 339-346 (1999)

\bibitem{Feldmann:1999uf}
T.~Feldmann,
Int. J. Mod. Phys. A \textbf{15}, 159-207 (2000)

\bibitem{Bramon:1997va}
A.~Bramon, R.~Escribano and M.~D.~Scadron,
Eur. Phys. J. C \textbf{7}, 271-278 (1999)

\bibitem{Cao:1999fs}
F.~G.~Cao and A.~I.~Signal,
Phys. Rev. D \textbf{60}, 114012 (1999)

\bibitem{Escribano:2005qq}
R.~Escribano and J.~M.~Frere,
JHEP \textbf{06}, 029 (2005)

\bibitem{Pham:2015ina}
T.~N.~Pham,
Phys. Rev. D \textbf{92}, 054021 (2015)

\bibitem{Michael:2013gka}
C.~Michael \textit{et al.} [ETM],
Phys. Rev. Lett. \textbf{111}, 181602 (2013)

\bibitem{Kroll:2005sd}
P.~Kroll,
Mod. Phys. Lett. A \textbf{20}, 2667-2684 (2005)

\bibitem{Cao:2012nj}
F.~G.~Cao,
Phys. Rev. D \textbf{85}, 057501 (2012)

\bibitem{LHCb:2025sgp}
R.~Aaij \textit{et al.} [LHCb],
JHEP \textbf{10}, 113 (2025)

\bibitem{CELLO:1990klc}
H.~J.~Behrend \textit{et al.} [CELLO],
Z. Phys. C \textbf{49}, 401-410 (1991)

\bibitem{CLEO:1997fho}
J.~Gronberg \textit{et al.} [CLEO],
Phys. Rev. D \textbf{57}, 33-54 (1998)

\bibitem{BaBar:2011nrp}
P.~del Amo Sanchez \textit{et al.} [BaBar],
Phys. Rev. D \textbf{84}, 052001 (2011)

\bibitem{BaBar:2006ash}
B.~Aubert \textit{et al.} [BaBar],
Phys. Rev. D \textbf{74}, 012002 (2006)

\bibitem{Choi:2017zxn}
H.~M.~Choi, H.~Y.~Ryu and C.~R.~Ji,
Phys. Rev. D \textbf{96}, no.5, 056008 (2017)

\bibitem{GomezDumm:2016bxp}
D.~Gomez Dumm, S.~Noguera and N.~N.~Scoccola,
Phys. Rev. D \textbf{95}, 054006 (2017)

\bibitem{Ding:2018xwy}
M.~Ding, K.~Raya, A.~Bashir, D.~Binosi, L.~Chang, M.~Chen and C.~D.~Roberts,
Phys. Rev. D \textbf{99}, 014014 (2019)


\bibitem{Hu:2026rfj}
D.~D.~Hu, X.~G.~Wu, Y.~J.~Zhang, H.~B.~Fu and T.~Zhong,
Phys. Rev. D \textbf{114}, 016003 (2026)

\bibitem{Holz:2024diw}
S.~Holz, M.~Hoferichter, B.~L.~Hoid and B.~Kubis,
JHEP \textbf{04}, 147 (2025)

\bibitem{Holz:2024lom}
S.~Holz, M.~Hoferichter, B.~L.~Hoid and B.~Kubis,
Phys. Rev. Lett. \textbf{134}, 171902 (2025)

\bibitem{Messerli:2025rnv}
N.~Messerli, M.~Hoferichter, B.~L.~Hoid, S.~Holz and B.~Kubis,
JHEP \textbf{04}, 088 (2026)

\bibitem{Choi:2007yu}
H.~M.~Choi and C.~R.~Ji,
Phys. Rev. D \textbf{75}, 034019 (2007)

\bibitem{Jaus:1999zv}
W.~Jaus,
Phys. Rev. D \textbf{60}, 054026 (1999)

\bibitem{Cheng:2003sm}
H.~Y.~Cheng, C.~K.~Chua and C.~W.~Hwang,
Phys. Rev. D \textbf{69}, 074025 (2004)

\bibitem{Choi:2013mda}
H.~M.~Choi and C.~R.~Ji,
Phys. Rev. D \textbf{89}, no.3, 033011 (2014)

\bibitem{Choi:2017uos}
H.~M.~Choi and C.~R.~Ji,
Phys. Rev. D \textbf{95}, no.5, 056002 (2017)

\bibitem{Chang:2018zjq}
Q.~Chang, X.~N.~Li, X.~Q.~Li, F.~Su and Y.~D.~Yang,
Phys. Rev. D \textbf{98}, no.11, 114018 (2018)

\bibitem{Bakamjian:1953kh}
B.~Bakamjian and L.~H.~Thomas,
Phys. Rev. \textbf{92}, 1300-1310 (1953)

\bibitem{Keister:1991sb}
B.~D.~Keister and W.~N.~Polyzou,
Adv. Nucl. Phys. \textbf{20}, 225-479 (1991)

\bibitem{Xu:2025ntz}
S.~Xu, X.~N.~Li and X.~G.~Wu,
Sci. China Phys. Mech. Astron. \textbf{69}, 231012 (2026)

\bibitem{Li:2026wmb}
X.~N.~Li, S.~Xu and Q.~Chang,
Phys. Rev. D \textbf{113}, 014018 (2026)

\bibitem{Li:2026wad}
X.~N.~Li, S.~Xu and Q.~Chang,
Phys. Lett. B \textbf{879}, 140646 (2026)

\bibitem{Xu:2026zli}
S.~Xu, X.~N.~Li, J.~Z.~Han, B.~H.~Cheng, L.~L.~Chen and Q.~Chang,
JHEP \textbf{05}, 138 (2026)

\bibitem{Lepage:1980fj}
G.~P.~Lepage and S.~J.~Brodsky,
Phys. Rev. D \textbf{22}, 2157 (1980)

\bibitem{Chernyak:1983ej}
V.~L.~Chernyak and A.~R.~Zhitnitsky,
Phys. Rept. \textbf{112}, 173 (1984)


\bibitem{Hu:2021zmy}
D.~D.~Hu, H.~B.~Fu, T.~Zhong, L.~Zeng, W.~Cheng and X.~G.~Wu,
Eur. Phys. J. C \textbf{82}, 12 (2022)

\bibitem{Kroll:2012gsh}
P.~Kroll and K.~Passek-Kumericki,
J. Phys. G \textbf{40}, 075005 (2013)

\bibitem{Agaev:2014wna}
S.~S.~Agaev, V.~M.~Braun, N.~Offen, F.~A.~Porkert and A.~Sch{\"a}fer,
Phys. Rev. D \textbf{90}, 074019 (2014)

\bibitem{Ball:2004ye}
P.~Ball and R.~Zwicky,
Phys. Rev. D \textbf{71}, 014015 (2005)

\bibitem{A2:2013wad}
P.~Aguar-Bartolom\'e \emph{et al.} [A2 Collaboration],
Phys. Rev. C \textbf{89}, 044608 (2014)


\bibitem{BESIII:2015zpz}
M.~Ablikim \textit{et al.} [BESIII],
Phys. Rev. D \textbf{92}, no.1, 012001 (2015)


\bibitem{ParticleDataGroup:2024cfk}
S.~Navas \textit{et al.} [Particle Data Group],
Phys. Rev. D \textbf{110}, 030001 (2024)


\end{thebibliography}
\end{document}